\documentstyle[12pt,epsf,wrapfig]{article}
\newcommand{\id}{\mbox{d}}
\newcommand{\GeV}{\mbox{GeV}}
\newcommand{\GeVs}{\mbox{GeV}^2}
\newcommand{\pN}{{\rm N}}

\newcommand{\beq}{\begin{equation}}
\newcommand{\eeq}{\end{equation}}
\newcommand{\beqa}{\begin{eqnarray}}
\newcommand{\eeqa}{\end{eqnarray}}
\newcommand{\bmat}{\left(\begin{array}}
\newcommand{\emat}{\end{array}\right)}
\newcommand{\mcenter}[2]{\hbox to #1cm{\hss$\displaystyle#2$\hss}}
\newcommand{\ignore}[1]{\relax}
\newcommand{\K}{{\rm K}}                   
\newcommand{\D}{{\rm D}}                   
\newcommand{\Dn}{{\rm D}^0{}}          

\newcommand{\uq}{{\rm u}}                  
\newcommand{\cq}{{\rm c}}                  
\newcommand{\gq}{{\rm g}}                  
\newcommand{\ccb}{{\rm c\bar{c}}}     
\newcommand{\G}{g}                         
\renewenvironment{thebibliography}[1]
        {\begin{list}{\arabic{enumi}.}
        {\usecounter{enumi}\setlength{\parsep}{0pt}
         \setlength{\itemsep}{0pt}
         \settowidth
        {\labelwidth}{#1.}\sloppy}}{\end{list}}
 
\begin{document}
\vbox to 0pt{\vss
\begin{flushright}
hep-ex/9611016
\end{flushright}}
\begin{center}
\def\thefootnote{\fnsymbol{footnote}}%
\makeatletter
\def\@makefnmark{\hbox to\z@{$\m@th^{\@thefnmark}$\hss}}%
\makeatother
{\large \bf
Spin Physics with COMPASS\footnote
{Presented at SPIN '96---The 12th.\ International Symposium on High-Energy
 Spin Physics (Amsterdam, Sept.\,1996).}}

\vspace{5mm}
G.K. Mallot\\
\vspace{5mm}
{\small\it
Institut f\"ur Kernphysik der Universit\"at Mainz, Becherweg 45,
D-55099 Mainz, Germany\\
\small
On Behalf of the COMPASS Collaboration
 }
\end{center}
\setcounter{footnote}{0}%

\begin{center}
%
\begin{minipage}{130 mm}
\small
The recently proposed COMPASS experiment at CERN attempts a measurement of 
the gluon polarisation with a precision of $\delta\Delta\G/\G\simeq0.1$. 
The experiment uses open charm muoproduction to tag the photon-gluon
fusion process.
\end{minipage}
\end{center}


\noindent
One of the most urgent questions in understanding the nucleon's spin structure
is the polarisation of gluons, $\Delta \G/\G$.
A large value of $\Delta \G$ could explain the smallness
of the contribution of the quark spins to the nucleon spin,
$\Delta\Sigma$ \cite{SMC95a,E143_95b}.
Hints for
a large value of $\Delta \G$ in the order of 2--3~$\hbar$ at $Q^2=10~\GeVs$
come from a recent QCD analysis \cite{BaF95b} of existing $g_1$ data.
However, an unambiguous determination of $\Delta \G$, can only be
obtained from a process involving the gluon distribution in leading order.
A particularly clean such process is open charm production via the 
photon-gluon fusion process, $\gamma \gq \rightarrow \ccb$, shown in Fig.~1.
Contributions from quark distributions can be neglected because there is no 
or only a small
intrinsic charm quark content in the nucleon. 
The scale is set by the charm quark mass, $4m_\cq\simeq10~\GeVs$.

\begin{wrapfigure}[16]{r}{8cm}
\mbox{\epsfxsize7cm\epsffile[50 0 585 500]{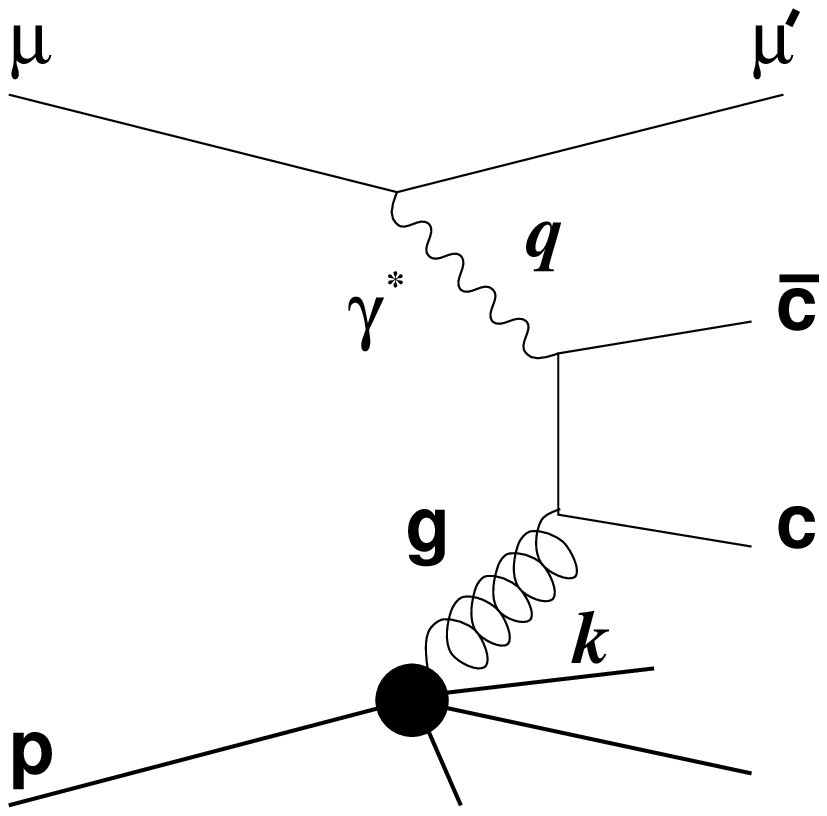}}

\small\noindent Figure 1:
The photon-gluon fusion diagram.
\label{fig:pgf_1}
\end{wrapfigure}
The main goal of the COMPASS experiment is the determination of
$\Delta \G/\G$ from the cross section asymmetry for polarised open charm 
muoproduction from a fixed polarised target. 
The tagging of charm events is based on the
identification of $\Dn$ mesons via their $\D\rightarrow\K\pi$ decay channels.
The experiment will also provide high statistics
data for $g_1$, semi-inclusive muon scattering, and the transversity
structure function, $h_1$.
The layout of the apparatus is similar to that of the SMC experiment and
parts of this setup will be used. 

\begin{figure}[t]
\begin{minipage}[t]{0.485\linewidth}
\begin{center}
\mbox{\epsfysize=0.9\hsize\epsffile[0 0 567 590]{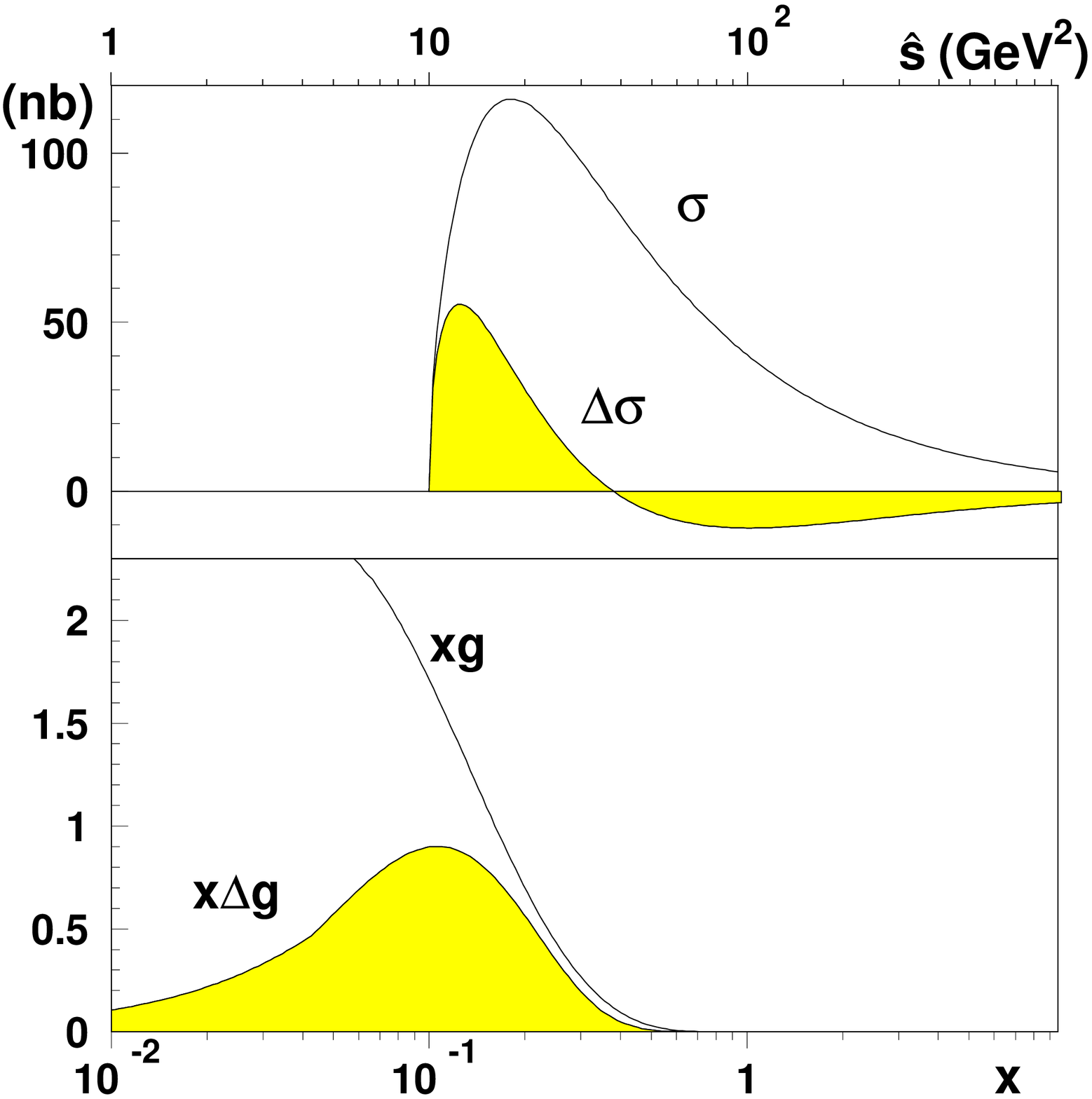}}
\end{center}

\small\noindent Figure 2:
The photon-gluon cross sections $\sigma$
and $\Delta\sigma$ as a function of $\hat s$ (top) and
$x\G$ and $x\Delta\G$, as a function of $x=x_\gq$ (bottom).
\label{fig:sigma}
\end{minipage}
\hfill
\begin{minipage}[t]{0.485\linewidth}
\begin{center}
\mbox{\epsfysize=0.9\hsize\epsffile{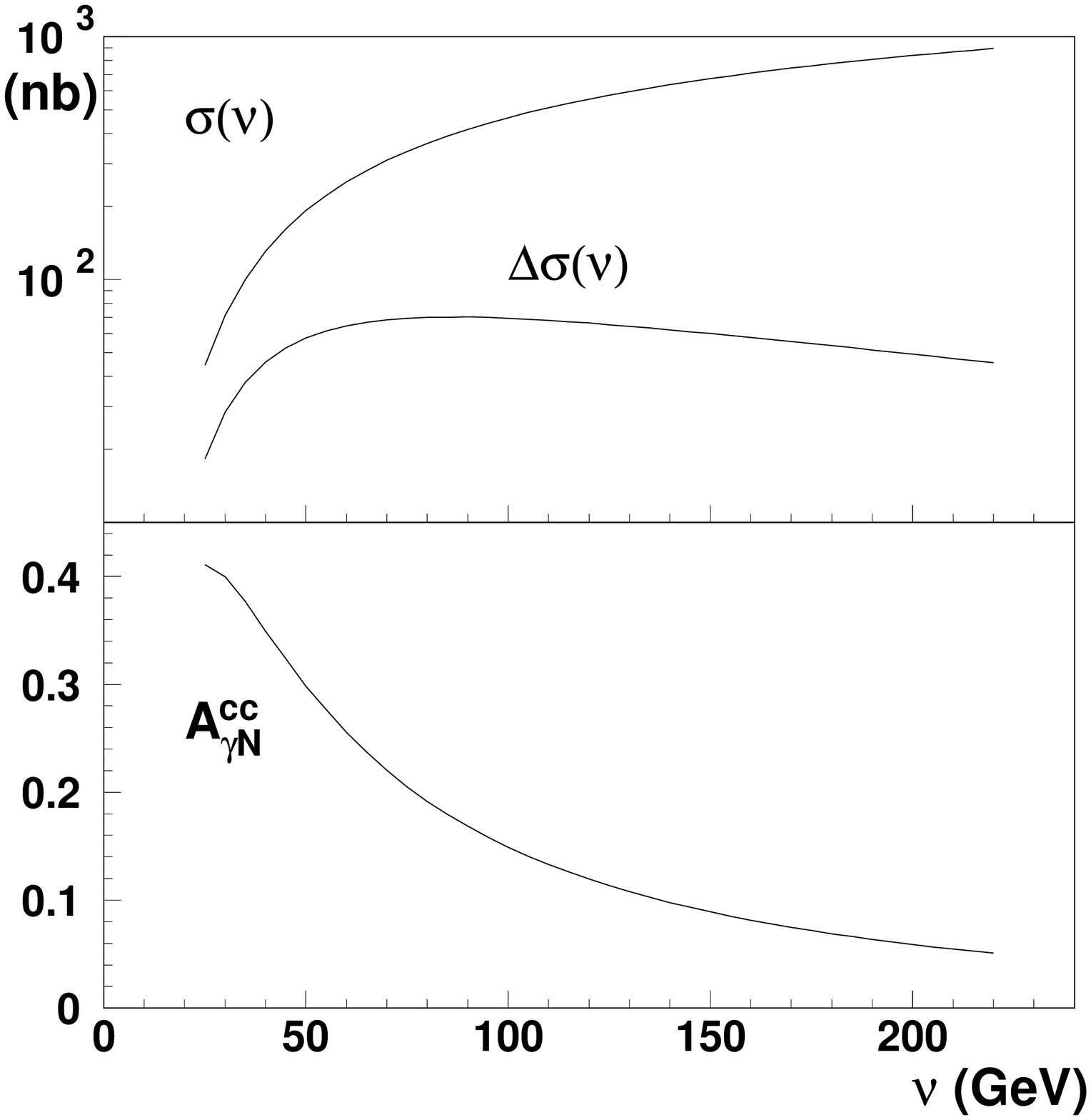}}
\end{center}

\small\noindent
Figure 3:
The photon-nucleon cross sections,
$\sigma$ and $\Delta\sigma$, (top) and
the $A^\ccb_{\gamma\pN}$, (bottom) as a
function of $\nu$.
\label{fig:gammaN}
\end{minipage}
\end{figure}
 

For real or quasi-real photons the charm-production cross section via the
photon-gluon fusion process can be written as
\begin{equation}
\sigma^{\gamma \gq\rightarrow\ccb} = \sigma(\hat{s})+\lambda_{\gamma}\lambda_\gq
\Delta\sigma(\hat{s}),
\end{equation}
where $\hat{s}=(q+k)^2$ is the energy squared and
$\lambda_{\gamma,\gq}$ are
the helicities in the photon-gluon c.m.\ system.
The spin-averaged and the spin-dependent
part, $\sigma(\hat{s})$ and $\Delta\sigma(\hat{s})$, are known to next-to-leading 
and leading order \cite{Wat82,GlR88}, respectively.
Both terms, $\sigma$ and $\Delta\sigma$, rise sharply at the threshold,
$\hat{s}=4m_\cq^2$, and $\Delta\sigma$ changes sign at about four times
the threshold (Fig.~2).
The photon-nucleon cross-section asymmetry, $A_{\gamma\pN}^{\ccb}$,
for the process $\gamma\pN\rightarrow\ccb$ is shown in Fig.~3. It is obtained by 
integrating the cross sections over the kinematically allowed range 
$x_\gq^{\rm min}\le x_\gq \le 1$
\begin{equation}
\label{eq:AgN}
A_{\gamma \pN}^{\ccb}(\nu)=
\frac{\Delta \sigma^{\gamma\pN\rightarrow \cq\bar{\cq}X}}
{\sigma^{\gamma\pN\rightarrow \cq\bar{\cq}X}} =
\frac{\int_{4m_\cq^2}^{2M\nu}\id\hat{s}\,\Delta
\sigma(\hat{s})\,\Delta \G(x_\gq,\hat{s})}
{\int_{4m_\cq^2}^{2M\nu}\id\hat{s}\,\sigma(\hat{s})\,\G(x_\gq,\hat{s})}.
\label{eq:acc}
\end{equation}
Here $x_\gq=\hat{s}/2M\nu$ denotes the nucleon-momentum fraction carried by
the gluon, which in this process is different from the kinematic variable,
$x=Q^2/2M\nu$.

From coherence arguments \cite{BrB94} it is expected that
the gluon polarisation behaves like $\Delta\G(x)/\G(x)\propto x$ for $x\rightarrow0$.
Such a behaviour is used in most parametrisations of the polarised gluon distribution
function.
Therefore, $\Delta g/g$ is expected to be large only at rather large values of
$x$. For $x_\gq\ge 0.1$ the gluon polarisation can exceed 0.5.
This value of $x_{\rm g}$ corresponds to a photon energy of order 50~GeV for
which the $\hat s$ and $x_\gq$ axes in Fig.~2 correspond to each other. 
For the rate estimates we use the polarised gluon distribution of 
Ref.~\cite{GeS94} (set~B) shown in Fig.~2.

COMPASS is a fixed-target experiment
similar to that of the SMC and
uses a
``\underline{Co}mmon \underline{M}uon and \underline{P}roton
\underline{A}pparatus for \underline{S}tructure and
\underline{S}pectroscopy''.
Apart from the spin physics programme discussed here in part
the proposal \cite{COMPASS96} also contains a spectroscopy programme with hadron
beams.
%
The experiment is statistics limited even with the
compared to the SMC experiment five times higher muon intensity of
$2\times10^8$ muons per spill of 2.4~s every 14.4~s.
The necessity to detect pions and kaons from the $\D$ decays in
a wide angular range
requires a two stage magnetic spectrometer with particle identfication in 
both spectrometer stages.
Downstream of the polarised target the new hadron stage of the spectrometer
covers hadron angles in the range $\pm200~$mrad. Its large-aperture dipole magnet 
provides a bending power of 1~Tm.
In the second stage of the spectrometer the scattered muon and fast hadrons
are measured. It uses the present SMC spectrometer magnet.
Particle identification will be performed by ring-imaging Cherenkov counters
and by electromagnetic and hadronic calorimeters in each of the two
spectrometer stages. The upstream and downstream RICH provide pion-kaon
separation in the momentum range 3--65~GeV/$c$ and 30--120~GeV/$c$, respectively.
Existing lead-glass arrays will be used
for the electromagnetic calorimeters. The hadronic calorimeters will in
the muon programme mainly serve trigger purposes and tag deep-inelastic
events. Both spectrometer stages end with a muon wall consisting of a
hadron absorber followed by tracking chambers and trigger hodoscopes.
The tracking in the beam region will be performed by scintillating fibre
detectors.
The large angles of the produced hadrons with respect to the incoming
beam also require a new solenoid for polarised target with an opening of
about $\pm 200$~mrad matching that of the first spectrometer stage.
As target materials lithium deuteride, $^6$LiD, for the deuteron
and ammonia, NH$_3$, for the proton are foreseen,
polarised to 50~\% and 85~\%, respectively.
The nuclear structure of $^6$LiD is well described by the
``alpha + deuteron'' picture, which results in the favourable
dilution factor of $f=0.50$ compared to $f=0.16$ for
ammonia.
The diameter of the two oppositely polarised 60~cm long target cells will
be reduced to 3~cm in order to minimise the amount of material traversed.
The nominal luminosity amounts to 
${\cal L} = 5\times10^{32}\mbox{cm}^{-2}\mbox{s}^{-1}$.
The minimum $x_\gq$ value one wants to access determines the maximum photon 
energy, $\nu$, needed.
The muon energy should only be slightly larger than $\nu$, in order to
obtain a large average depolarisation factor, $D$.
A muon energy of 100~GeV appears to be the optimal choice for $x_\gq\simeq0.1$.
The muon energy can be increased up to 200~GeV to explore smaller values of
$x_\gq$.

For the kinematics of the COMPASS experiment we find in average 1.2 $\Dn$
mesons per initial $\ccb$ pair including
$\D$ mesons from $\D^\star$ decays.
We concentrate on the two-body decays
$     \Dn (\cq\bar{\uq})\rightarrow\K^-\pi^+$ and $c.c.$
with branching ratios of 4~\%.
A major concern is the combinatorial background of kaon-pion pairs
within the $\D$ mass window.
The mass of the $\D$ meson will be recontructed with a resolution of
$\sigma_{M_\D}\simeq10$~MeV.
In the range $25<\nu<85$~GeV the open charm production cross section 
amounts to 2~nb compared to 500~nb for ordinary photoproduction.
Often the distance between the production and the decay vertex is used
in charm experiments to clean up the kaon-pion sample.
This technique cannot be applied in the COMPASS experiment, because
this distance of a few mm cannot be resolved due to multiple scattering in
the target.
Kaons emitted at large angles, $\theta_\K^*$, in
the $\D$'s rest frame with respect to the $\D$'s direction of flight in 
the laboratory frame have large transverse momenta.
On the other hand,
kaons from ordinary fragmentation have small transverse momenta
and thus dominantly mimic decays with small $\theta_\K^*$.
The background rejection was studied in Monte Carlo simulations using the 
AROMA \cite{AROMA_2.1} and JETSET event generators for the photon-gluon fusion
process and the background, respectively (Fig.~4). 
The best result is obtained with the requirements
$|\cos\theta_\K^*|\le0.5$ and $z_\D=E_\D/\nu\ge0.25$, which improve the
background-to-signal ratio by a factor 1750 to about
$N^{\rm B}/N^\ccb\simeq3.8$
on the expense of
loosing 65~\% of the $\D$ mesons.
\begin{figure}
 \begin{center}
  \mbox{\epsfxsize=0.9\hsize\epsffile{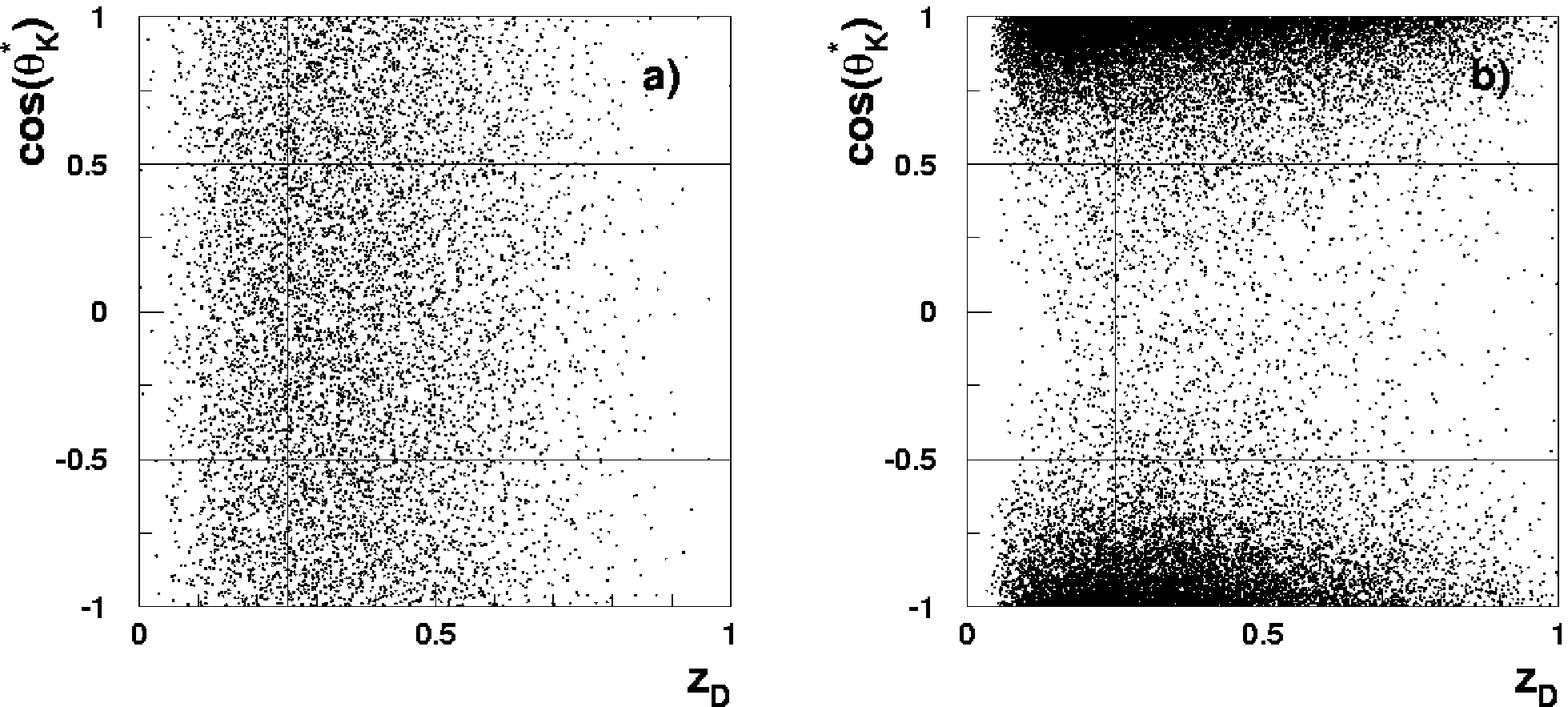}}

\end{center}
\small\noindent Figure 4:
Monte-Carlo simulation: a) $\D\rightarrow\K\pi$  and b) combinatorial
background versus $z_\D$ and $|\cos\theta^\star_\K|$.
The lines indicates the cuts used in the rate estimates.
\label{fig:d0zc}
\end{figure}
%
For a running time of 2 1/2 years with 150 days/year and assuming a combined
efficiency of 0.25 for the muon beam and the experimental apparatus the statistical
error of the measured asymmetry is $\delta A^\ccb_{\gamma N}=0.076$.
Due to the higher figure of merit of the $^6$LiD target a similar precision
will already be reached after the first 1 1/2 years.
The result can be improved using $\D^\star$ tagging by the soft pion from the 
decay
$
\D^{\star+}\rightarrow\Dn\pi^+_s\rightarrow(\K^-\pi^+)\pi^+_s.
$
Considering only the soft pions with momenta larger than 1~GeV/$c$
and taking possible re-interaction in the target
into account the statistical error of the asymmetry reduces to
\beq
\delta A^\ccb_{\gamma N}=0.05
\hskip 1cm \mbox{corresponding to} \hskip 0.5cm
\delta\frac{\Delta g}{g}=0.14.
\eeq
As in the inclusive case the muon-nucleon asymmetry is reduced from the 
virtual-photon asymmetry by the depolarisation factor,
$A^{\ccb}_{\mu\pN}=DA_{\gamma\pN}^\ccb$ (Fig.~5).
The sensitivity to $\Delta g/g$ peaks at $x_\gq=0.14$ and covers the
range $0.07\le x_\gq \le 0.4$ (Fig.~6).
Apart from the c.m.\ energy, $\hat s$,
the asymmetry for the elementary photon-gluon fusion process,
$\Delta\sigma(\hat s,\hat \theta)/\sigma(\hat s)$, also depends on
on the c.m.\ angle,
$\hat \theta$, between the photon-gluon axis and the $\ccb$
axis. 
The sensitivity is larger for small angles, $\hat\theta$, 
corresponding in the laboratory
frame to small transverse momenta, $p_T$.
Rejecting $\D$ mesons with $p_T>1~\GeV/c$ thus yields a larger analysing power
leading to
\beq
\delta A^\ccb_{\gamma N}=0.04
\hskip 1cm \mbox{corresponding to} \hskip 0.5cm
\delta\frac{\Delta g}{g}=0.11.
\eeq
The three and four-body decay channels
may further improve the precision, in particular if the $\D^\star$
tagging can be applied. This is still under investigation.

\begin{figure}
   \begin{minipage}[t]{0.485\linewidth}
      \begin{center}
         \mbox{\epsfxsize=0.7\hsize\epsffile[50 0 500 500]{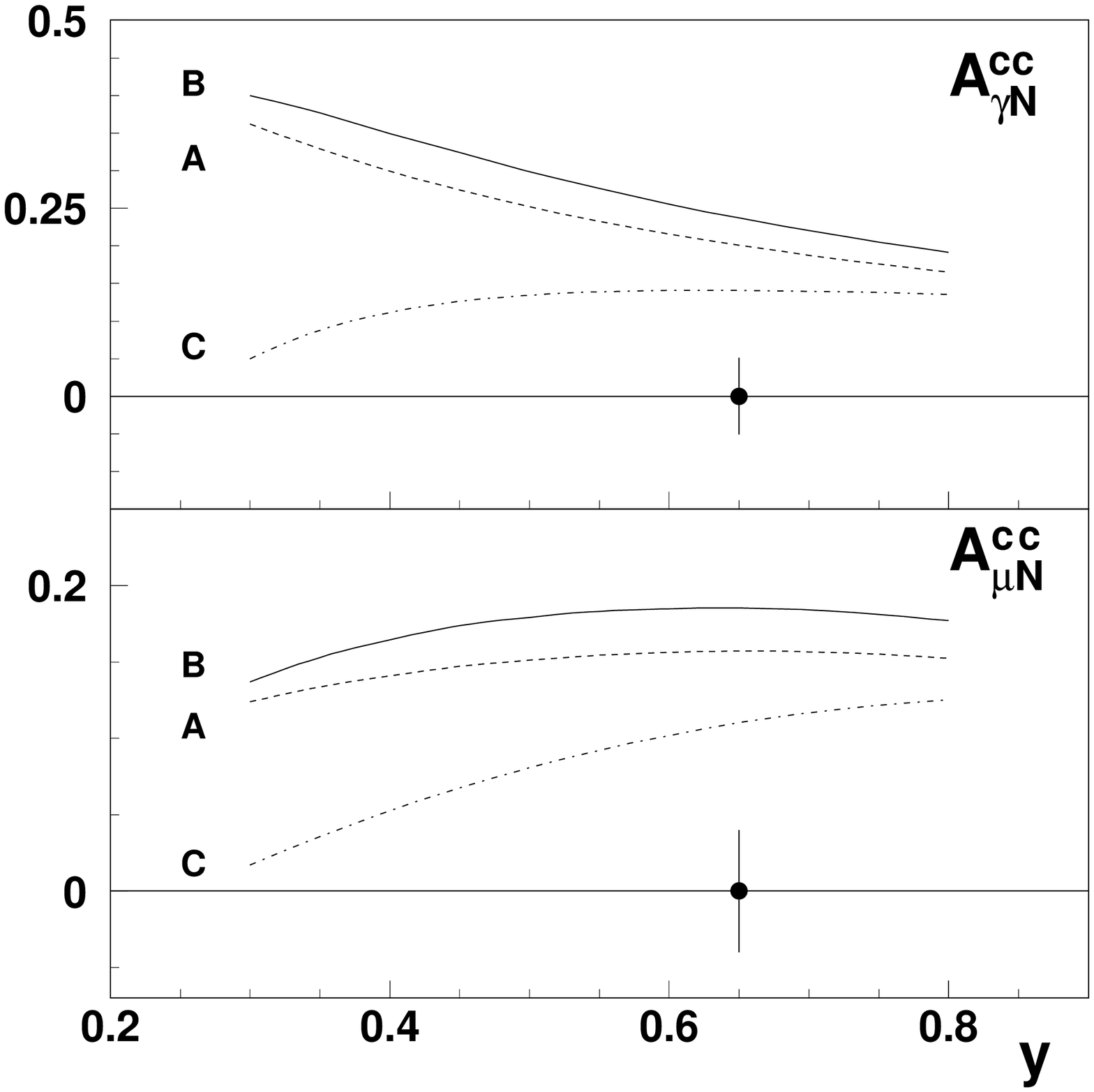}}
         \end{center}
\small\noindent Figure 5:
               Photon-nucleon asymmetry (top),
               and muon-nucleon asymmetry (bottom),
               as a function of $y=\nu/E_\mu$ for the three
               gluon distributions of Ref.~\protect\cite{GeS94}.
               The data points indicate the projected precision of the 
               \mbox{COMPASS} measurement.
      \label{fig:casy}
      \end{minipage}
   \hfill
   \begin{minipage}[t]{0.485\linewidth}
      \begin{center}
         \mbox{\epsfxsize=\hsize\epsffile[ 0 20 283 255]{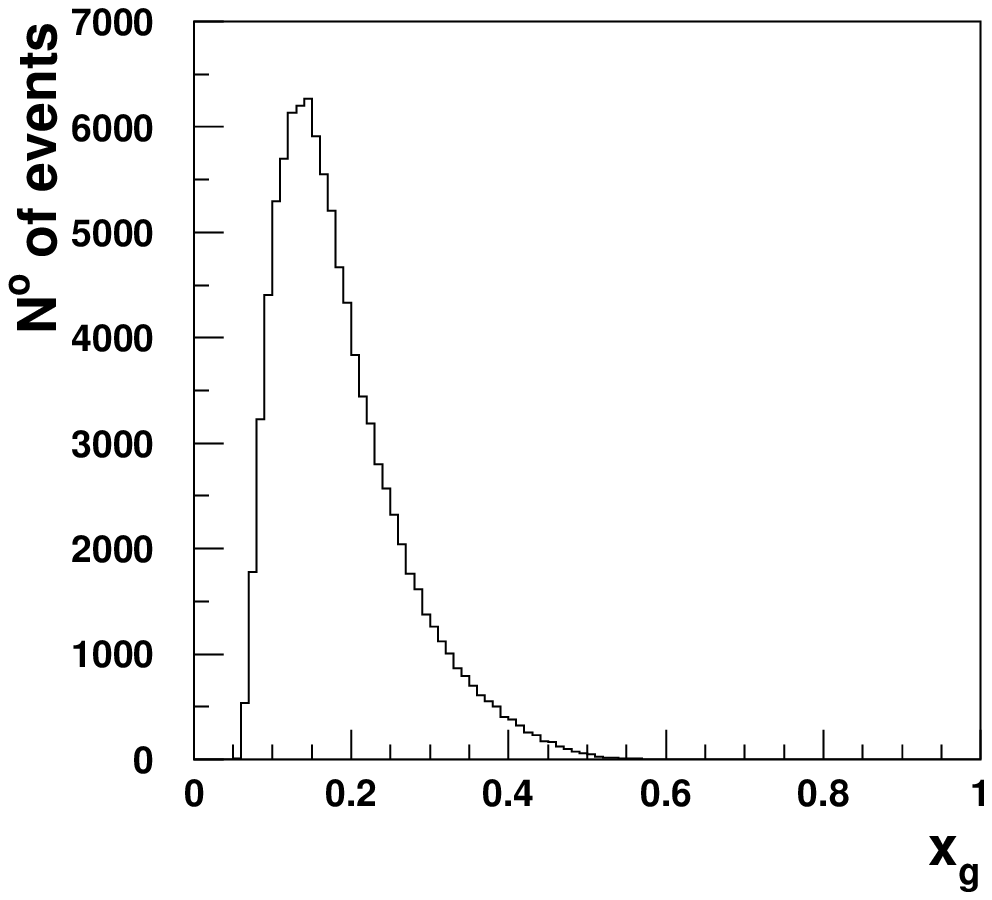}}
         \end{center}
\small\noindent Figure 6:
               Number of accepted photon-gluon fusion events
               as a function of $x_\gq$ for the range
               $35\le \nu \le 85~\GeV$.
      \label{fig:xg}
      \end{minipage}
   \end{figure}

Apart from the measurement of $\Delta \G/\G$ COMPASS offers 
a rich spin-physics programme at high $Q^2$ with a high luminosity including
the transversity structure function $h_1$,
spin-flavour decomposition of the structure functions, and
lambda polarisation in both the target and current fragentation regions.
The SPSLC has recommended the COMPASS experiment for approval. After
commissioning in 1999 data taking could start in the year 2000.

I like to thank the COMPASS Collaboration and in particular my Mainz colleagues,
A. Bravar, D. von Harrach, E.~Kabu\ss, and J.~Pretz 
for the help in preparing this manuscript.

\end{document}